\documentclass[twocolumn,aps,prc,superscriptaddress,showpacs,nobibnotes]{revtex4}
\begin{document}
\draft
\title{Strangeness  $s = -6 $ dibaryon }

\author{DAI Lian-Rong} \thanks{E-mail: dailr@lnnu.ac.cn}
 \affiliation{Department of Physics,Liaoning Normal University, 116029, Dalian, P. R. China}
\author{ZHANG Zong-Ye}\thanks{E-mail: zhangzy@mail.ihep.ac.cn}
\author{YU You-Wen}
\affiliation{Institute of High Energy Physics, 100039, Beijing, P.
R. China}

\begin{abstract} \textnormal{\small {
The structure of $(\Omega\Omega)_{0^+}$  dibaryon  with strangeness
$s=-6$  is studied in the extended chiral $SU(3)$ quark model, in
which vector meson exchange  dominates the short range interaction.
The resonating group method (RGM) is adopted, in which the
$\Omega\Omega$ and $CC$ (hidden color) channels are involved. The
color screening effect and the effects of mixing of scalar mesons on
$(\Omega\Omega)_{0^+}$ are also investigated. }}
\end{abstract}



\maketitle

\section{Introduction}

\begin{table*}
\caption{Model parameters and the corresponding binding energies of
Deuteron. Meson masses and cutoff masses: $m_{\pi}=138$MeV,
$m_{K}=495$MeV, $m_{\eta}=548$MeV, $m_{\eta'}=958$MeV, $m_{\sigma'}=
m_{\epsilon}= 980$MeV, $m_{\kappa}=1430$MeV, $m_{\rho}=770$ MeV,
$m_{K^*}=892$MeV, $m_{\omega}=782$MeV, $m_{\phi}=1020$MeV,
$\Lambda=1500$MeV for $\kappa$ and $\Lambda=1100$MeV for other
mesons.}
\begin{small}
\begin {center}
\begin{tabular}{cccc}
\hline\hline
                       & Chiral $SU(3)$ quark model &
\multicolumn{2}{c}{~Extended  chiral $SU(3)$ quark model}   \\
                       &          & ~~~~~~ set I ~~~~~&  set
                        II
              \\ \hline
$b_u$ (fm)             & 0.5      & ~~~~~~ 0.45  ~~~~   & 0.45      \\
$g_{NN\pi}$            & 13.67    & 13.67    & 13.67     \\
$g_{ch}$               & 2.621    & 2.621    & 2.621     \\
$g_{chv}$              & 0        & 2.351    & 1.972     \\
$f_{chv}/g_{chv}$      & 0        & 0        & 2/3       \\
$m_{\sigma}$ (MeV)      & 595      & 535      & 547       \\
$g_u^2$     & 0.785    & 0.086    & 0.159     \\
$g_s^2$     & 0.571    & 0.003    & 0.062     \\
$a_{uu}$ (MeV/${\rm fm}^2$)     & 48.1     & 48.0     & 42.9      \\
$a_{us}$ (MeV/${\rm fm}^2$)     & 63.7     & 85.3     & 78.9      \\
$a^{0}_{uu}$ (MeV/${\rm fm}^2$)     & -43.6     & -74.4     & -65.3  \\
$a^{0}_{us}$ (MeV/${\rm fm}^2$)     & -50.8    & -100.2     & -89.9   \\
$B_{deu}$ (MeV)         & 2.13     & 2.19     & 2.14      \\
\hline\hline
\end{tabular}
\end{center}
\end{small}
\end{table*}

Since the H dibaryon was theoretically predicted ~\cite{Jaffe} by
Jaffe in 1977, searching dibaryons both theoretically and
experimentally has attracted worldwide attention. Various quark
models were proposed to study their possible existence. Different
from the deuteron, whose property can be well explained by meson
exchange mechanism on baryon level, dibaryon is supposed to be a
color singlet multi-quark system with a sufficiently smaller size
and dominance in quark-gluon degrees of freedom. Through these
studies, we have enriched our knowledge of the strong interaction
between quarks in the short -range region and some aspects about the
basic theory of strong interactions, Quantum Chromodynamics (QCD),
especially its nonperturbative effect.

Because of complexity in QCD nonperturbative effect at
lower-energy region, one has to develop QCD-inspired models. The
constituent quark model \cite{oka,Faessler1,Faessler2,zhang85} has
been quite successful in understanding hadron spectroscopy and the
short range behavior of hadron interactions even though we have
not been able to derive the constituent quark model directly from
QCD. Motivated by its success and combined with the chiral field
theory, the model was then modified as the chiral quark model
\cite{fernandez}. Later on, as a generalization of the $SU(2)$
linear $\sigma$ model, a chiral $SU(3)$ quark model was developed
to describe systems with strangeness \cite{zhang97}. The chiral
$SU(3)$ quark model has been quite successful in reproducing
energies of baryon ground states, binding energy of deuteron, the
nucleon-nucleon (NN) scattering phase shifts, and the
hyperon-nucleon (YN) cross sections by performing the resonating
group method (RGM) calculations.  Based on this model, Zhang {\it
et al} \cite{zhprc61} predicted that the $(\Omega\Omega)_{0^{+}}$
dibaryon with $s = -6 $ is the most possible dibaryon candidate
since its binding energy is large enough with small mean squared
root radius in the model calculations. A suggestion to search its
possible existence in heavy ion collision was also
made.\cite{zhprc61}



In Ref.\cite{dai}, we extended our chiral $SU(3)$ quark model to
include the coupling  between the quark and vector chiral fields.
Such extension was made mainly based on the following facts.
Firstly, in the study of NN interactions on quark level, the
short-range feature can be explained by one gluon exchange (OGE)
interaction and the quark exchange effect, while in the traditional
one boson exchange (OBE) model on baryon level it comes from vector
meson ($\rho,\omega, K^*$ and $\phi$) exchanges. Secondly, Glozman
and Riska proposed the boson exchange model \cite{glozman}, and
found that the OGE can be replaced by vector-meson coupling in order
to elucidate baryon structure. However, Isgur gave a critique on the
boson exchange model and insisted that OGE govern the baryon
structure \cite{isgur}. Furthermore, though a valence lattice QCD
result shown by Liu {\it et al} did support the Goldstone boson
exchange picture \cite{liu}, Isgur considered such conclusion
unjustified. Therefore, whether OGE or vector meson exchange is the
right mechanism in describing the short range part of the strong
interactions, or whether both of them are all important, is still a
challenge. Nevertheless, our study on deuteron structure and the
$NN$ scattering phase shifts in the extended chiral $SU(3)$ quark
model \cite{dai} did show that quark-vector chiral field coupling
interactions can substitute the OGE mechanism on quark level. In the
extended chiral $SU(3)$ quark model, instead of the OGE interaction,
the vector meson exchanges play a dominate role in the short range
part of the quark-quark interactions. Since geometric size of a
dibaryon is small, the short range feature of the interactions
should be important in describing its structure. Hence, a further
investigation on structure of $(\Omega\Omega)_{0^{+}}$ in the
extended chiral $SU(3)$ quark model should be helpful in resolving
this issue.

 In Refs. \cite{yuan,dai05}, the CC (hidden color) channel
and the color screening effect were investigated for
$(\Delta\Delta)_{3^+}$ (d*) dibaryon. In Refs. \cite{huang,huang1},
the chiral  SU(3) quark model and the extended  chiral SU(3) quark
model were used to study the baryon-meson interaction, in which
different mixing of scalar mesons $\sigma$ and $\epsilon$ were
discussed. It is interesting and important to study different mixing
of scalar mesons  for dibaryon system.

In this work, we will investigate structure of
$(\Omega\Omega)_{0^+}$ dibaryon in the extended chiral $SU(3)$ quark
model framework. A detailed analysis of the vector meson exchange,
the CC channel, the color screening effect, and variance of the
mixing angle between scalar mesons will be made. The paper is
arranged as follows. A brief introduction of the extended chiral
$SU(3)$ quark model is shown in section II. The results of the
binding energies of $(\Omega\Omega)_{0^+}$  in the extended chiral
$SU(3)$ quark model with discussions are provided in section III. A
summary is given in section IV.

\section{Formulation}
\subsection{The Model}
In the extended chiral $SU(3)$ quark model\cite{dai}, besides the
nonet pseudo-scalar meson fields and the nonet scalar meson fields,
the coupling between vector meson fields and quarks is also
considered. With this generalization, the Hamiltonian of the system
can be written as
\begin{eqnarray}
& H & =\sum\limits_{i}T_i-T_{\rm G}+\sum\limits_{i<j}V_{ij},
\end{eqnarray}
and
\begin{eqnarray}
& V_{ij} & =V_{ij}^{\rm conf}+V_{ij}^{\rm OGE}+V_{ij}^{\rm ch},
\end{eqnarray}
\begin{eqnarray}
& V_{ij}^{\rm ch} & = \sum^{8}_{a=0} V^{{\rm s}_a} (\vec{r}_{ij}) +
\sum^{8}_{a=0}
   V^{{\rm ps}_a} (\vec{r}_{ij})+ \sum^{8}_{a=0}V^{{\rm v}_a}
   (\vec{r}_{ij})
\end{eqnarray}
where $\sum\limits_{i}T_i-T_{\rm G}$ is the kinetic energy of the
system, and $V_{ij}$ includes all the interactions between two
quarks. $V_{ij}^{\rm conf}$ is the confinement potential taken in
quadratic form, $V_{ij}^{\rm OGE}$ is the OGE interaction, and
$V_{ij}^{\rm ch}$ represents the interactions from the chiral field
coupling, which, in the extended chiral $SU(3)$ quark model,
includes the scalar meson exchange $V_{ij}^{\rm s}$, the
pseudo-scalar meson exchange $V_{ij}^{\rm ps}$ and the vector meson
exchange $V_{ij}^{\rm v}$ potentials. In Eq.(2), the OGE is taken in
the usual form \cite{dai}, while the confinement potential is chosen
in the quadratic form
\begin{eqnarray}
V_{ij}^{\rm
conf}=-\lambda_{i}^{c}\cdot\lambda_{j}^{c}~a_{ij}^{0}-
\lambda_{i}^{c}\cdot\lambda_{j}^{c}a_{ij}
r_{ij}^2~.
\end{eqnarray}
The quark-chiral field (scalar,pseudo-scalar and vector nonet
mesons) induced interactions are
\begin{eqnarray}
V^{{\rm s}_a} (\vec{r}_{ij}) & = & -C(g_{\rm ch}, m_{s_a},
\Lambda_{c}) X_{1}(m_{s_a}, \Lambda_{c}, r_{ij})
\lambda_{a}(i)\lambda_a (j)\nonumber\\
 &&{} + V^{\vec{\ell} \cdot \vec{s}}_{s_a} (\vec{r}_{ij})\,\, ,
\end{eqnarray}
\begin{eqnarray}
V^{{\rm ps}_a}(\vec{r}_{ij}) & = & C(g_{\rm ch}, m_{{\rm ps}_a},
\Lambda_{c}) \frac{m_{{\rm ps}_a}^{2}}{12m_{qi}m_{qj}}\{
X_{2}(m_{{\rm ps}_a}, \Lambda_{c},
r_{ij})\nonumber\\
 &&{}
 (\vec{\sigma}_{i} \cdot \vec{\sigma}_{j})
  + \left ( H(m_{{\rm ps}_a} r_{ij}) - (\frac{\Lambda_{c}}{m_{{\rm ps}_a}} )^{3}
H(\Lambda_{c} r_{ij} ) \right )\nonumber\\
 &&{}
\hat{S}_{ij}\} \lambda_{a}(i) \lambda_{a}(j)\,\, ,
\end{eqnarray}
\begin{eqnarray}
 V^{{\rm v}_a} (\vec{r}_{ij}) & = & C(g_{\rm chv}, m_{v_a}, \Lambda_c)
X_{1}(m_{v_a}, \Lambda_c, r_{ij})\lambda_{a}(i)\lambda_a(j) \nonumber\\
  & + & C(g_{\rm chv}, m_{v_a}, \Lambda_c)
\frac{m_{v_a}^{2}}{6m_{qi}m_{qj}}(1+\frac{f_{\rm chv}}{g_{\rm
chv}} \frac{m_{qi}+m_{qj}}{M_p}\nonumber\\
 &&{}
+\frac{f_{\rm chv}^2}{g_{\rm chv}^2}\frac{m_{qi} m_{qj}}{M_p^2})
 \{  X_{2}(m_{v_a}, \Lambda_c,r_{ij})
(\vec{\sigma}_{i} \cdot \vec{\sigma}_{j})  \nonumber\\
  & - &\frac{1}{2} \left ( H(m_{v_a} r_{ij}) - (\frac{\Lambda_c}{m_{v_a}}
)^{3} H(\Lambda_c r_{ij} ) \right )\hat{S}_{ij} \}\nonumber\\
 &&{}
\lambda_{a}(i)\lambda_{a}(j) +  V^{\vec{\ell} \cdot \vec{s}}_{{\rm
v}_a} (\vec{r}_{ij}),
\end{eqnarray}
\begin{eqnarray}
  V^{\vec{\ell} \cdot \vec{s}}_{s_{a}} (\vec{r}_{ij}) & = & -C(g_{ch},
        m_{s_{a}}, \Lambda_{c})
        \frac{m^{2}_{s_{a}}}{4m_{qi} m_{qj}}
       \{ G (m_{s_{a}} r_{ij} )  \nonumber\\
  & - & (\frac{\Lambda_{c}}{m_{s_{a}}} )^{3}
G(\Lambda_{c} r_{ij})\}\vec{L} \cdot (\vec{\sigma}_{i} +
\vec{\sigma}_{j} )
  \lambda_{a}(i)\lambda_{a}(j),~~~
\end{eqnarray}
and
\begin{eqnarray}
V^{\vec{\ell} \cdot \vec{s}}_{v_{a}} (\vec{r}_{ij})& = &
-C(g_{chv},m_{v_{a}}, \Lambda_{c})
        \frac{3m^{2}_{v_{a}}}{4m_{qi} m_{qj}}\nonumber\\
         &&{}
\left (1+\frac{f_{chv}}{g_{chv}}\frac{2(m_{qi}+m_{qj})}{3
M_p}\right) \times \{
G (m_{v_{a}} r_{ij} ) -\nonumber\\
 &&{}(\frac{\Lambda_{c}}{m_{v_{a}}}
)^{3}G(\Lambda_{c} r_{ij})\} \vec{L} \cdot (\vec{\sigma}_{i} +
\vec{\sigma}_{j})
  \lambda_{a}(i)\lambda_{a}(j),~~
\end{eqnarray}
respectively, where
\begin{eqnarray}
C(g, m, \Lambda)& = & \frac{g^{2}}{4\pi}
          \frac{\Lambda^{2} m }{\Lambda^{2} - m^{2} },\\
X_1(m, \Lambda, r)& = & Y(m r) -\frac{\Lambda}{m}Y(\Lambda r),\\
X_2(m, \Lambda, r)& = & Y(m r) -(\frac{\Lambda}{m})^3 Y(\Lambda r),\\
Y(x) & = & \frac{1}{x}e^{-x},\\
H(x) & = & (1+\frac{3}{x}+\frac{3}{x^2}) Y(x),\\
G(x) & = & \frac{1}{x}(1+\frac{1}{x}) Y(x),
\end{eqnarray}
\begin{eqnarray}
\hat{S}_{ij} = 3(\vec{\sigma}_{i} \cdot \hat{r})
        (\vec{\sigma}_{j} \cdot \hat{r})
      -(\vec{\sigma}_{i} \cdot \vec{\sigma}_{j}) ,
\end{eqnarray}
and $M_p$ is a mass scale, taken as proton mass.

\subsection{Determination of parameters}
In the model, $g_{ch}$ is the coupling constant for scalar and
pseudo-scalar chiral field coupling, which is determined according
to the following relation:
\begin{eqnarray}
\frac{g_{NN\pi}^2}{4\pi} = \frac{9}{25}~ \frac{m_u^2}{M_N^2}~
\frac{g_{ch}^2}{4\pi}~,
\end{eqnarray}
and ${g_{NN\pi}^2}/{4\pi}$ is taken to be the experimental value,
which is about $14$. $g_{chv}$ and  $f_{chv}$ are the coupling
constants for vector coupling and tensor coupling of the vector
meson field respectively.  The meson masses $m_{ps_{a}}$,
$m_{s_{a}}$, and $m_{v_{a}}$ are taken to be the corresponding
experimental values. Only the $m_{\sigma}$ is treated as an
adjustable parameter. According to the vacuum spontaneously breaking
theory, its value should satisfy the following relation
\cite{scadron}:
\begin{eqnarray}
m_{\sigma}^2= (2 m_u)^2 + m_{\pi}^2,
\end{eqnarray}
which can be regarded as almost reasonable when the value of
$m_{\sigma}$ is located in the range of $550 \sim 650$ MeV. In this
work, values of $g_{chv}$, $f_{chv}$, and $m_{\sigma}$ are taken to
be the same as Ref. \cite{dai}, which was  used in the study of $NN$
phase shift by fitting the binding energy of deuteron  . In our
calculation,  $\eta$ and $\eta'$ mesons are mixed from flavor
singlet $\eta_{0}$ and flavor octet $\eta_{8}$ mesons with
\begin{eqnarray}
\eta'=\eta_{8}\sin\theta^{PS}+\eta_{0}\cos\theta^{PS},\nonumber\\
\eta ~=\eta_{8}\cos\theta^{PS}-\eta_{0}\sin\theta^{PS},
\end{eqnarray}
in which the mixing angle is taken to be the experimental value with
$\theta^{PS}=-23^\circ$.  Other model parameters in the calculation
are fixed by the mass splitting among $N$, $\Delta$, $\Lambda$,
$\Sigma$, and $\Xi$, and the stability conditions of the octet
$(S=1/2)$ and decuplet $(S=3/2)$ baryons, which are all listed in
Table I.

\section{results and Discussion}
In Ref. \cite{zhprc61}, Zhang {\it et al} predicted the new
dibaryon candidate $(\Omega\Omega)_{0^+}$, which is a deeply bound
state formed due to the quark exchange effect and the chiral-quark
coupling and the fact that the quark exchange effect between two
$\Omega$ clusters makes these two $\Omega$ closed up, because
symmetry property of $(\Omega\Omega)_{0^+}$ is very special
\cite{zhprc61}. There are only six baryon-baryon systems with this
kind symmetry property, they are
$(\Delta\Delta)_{ST=30},(\Delta\Delta)_{ST=03},$
$(\Delta\Sigma^*)_{ST=3(1/2)},(\Delta\Sigma^*)_{ST=0(5/2)},$
$(\Xi^*\Omega)_{ST=0(1/2)}$ and $(\Omega\Omega)_{ST=00}$. Since
only $(\Omega\Omega)_{ST=00}$ can not decay through strong
interations among these six cases, in this sense
$(\Omega\Omega)_{ST=00}$ should be the most interesting dibaryon
candidate. In this section, we will further investigate the
structure of $(\Omega\Omega)_{0^+}$ dibaryon with respect to the
vector meson exchange, the CC channel, the color screening effect,
and the different mixing of scalar mesons.

\subsection{the effect from vector mesons exchange}

Structure of $(\Omega\Omega)_{0^+}$ is studied in the extended
chiral $SU(3)$ quark model in which the vector meson exchanges are
included. Our calculated results are shown in Table II, from which
it can be seen that the effect from the vector meson fields is quite
similar to that from the one gluon exchange (OGE) interaction. In
the extended chiral $SU(3)$ quark model, $(\Omega\Omega)_{0^+}$ is
still a deeply bound state with $135 \sim 158$ MeV in binding energy
when the vector meson exchanges control the short range part of the
quark-quark interaction. The detailed analysis can be found in Ref.
\cite{zhangctp}.
\begin{table}
\caption{Binding energy $B$ and rms $\overline{R}$ of the
$(\Omega\Omega)_{0^+}$ dibaryon in the coupled channel calculation.
$B=2M_{\Omega}-E_{(\Omega\Omega)_{0^+}}$,
$\overline{R}=\sqrt{\langle\,r^2 \rangle}$.}
\begin{small}
\begin {center}
\begin{tabular}{lllcc}
\hline\hline &&$B$(MeV)&$\overline{R}$(fm)\\
\hline
Chiral SU(3)  quark model&&170.9&0.62\\
&&\\
Extended chiral &~ set I~ &135.6&0.60\\
SU(3) quark model&~ set II~ &158.0&0.59\\

\hline\hline
\end{tabular}
\end{center}
\end{small}
\end{table}

\subsection{the effect from hidden color channel}
In Ref.\cite{yuan}, the CC (hidden color) channel was investigated
for the d* dibaryon. The mixture of the $L=0$ and $L=2$ states shows
that the effects of the tensor forces in the OGE and chiral-quark
field coupling were also considered. Their results show that the
coupled channel effect is much stronger than the $L$ state mixing
effect caused by the tensor interaction. It is interesting to see
what is the CC channel effect on the $(\Omega\Omega)_{0^+}$
dibaryon. Hence, the hidden color channel will be considered with
respect to the $(\Omega\Omega)_{0^+}$ dibaryon. According to
Harvey's work \cite{Harvey}, the hidden color channel wave function
can be constructed as
\begin{eqnarray}
|CC\rangle_{str=-6,ST=00}=~~~~~~~~~~~~~~~~~~~~~~~~~~~~~~~\nonumber\\
-\frac{1}{2}|\Omega\Omega\rangle_{ST=00} +\frac{5}{2}{\cal
A}_{SFC}|\Omega\Omega\rangle_{ST=00}~~
\end{eqnarray}
where $\cal{A}_{SFC}$ stands for the antisymmetrizer in the
spin-flavor-color space. The corresponding matrix elements of
spin-flavor-color operators of coupled channels  are given in
Table III.

After performing an off-shell transformation \cite{shen}, the
results are tabulated in Table IV in the coupled-channel
calculation. It is shown that the hidden color channel coupling only
has very little effect on the binding energy of
$(\Omega\Omega)_{0^+}$. Comparing with the study of d* dibaryon, in
Refs. \cite{yuan,dai05}, the results of the d* coupled-channel
calculation show that the hidden color channel coupling has obvious
effect on d* dibaryon. The binding energy of d* can be enhanced from
17 MeV to 23 MeV after the hidden color channel is included.

We make an analysis to see why the hidden color channel effect is
so different for the case of $(\Omega\Omega)_{0^+}$ and d*. From
the matrix elements of the Hamiltonian in the generator coordinate
method (GCM) calculation \cite{tang}, which can describe the
interaction between two clusters qualitatively, one can see that
the energy of the hidden color state $|CC\rangle_{str=-6,ST=00}$
is much higher than that of $(\Omega\Omega)_{0^+}$ state, and the
cross matrix elements between $(\Omega\Omega)_{0^+}$ and its
corresponding hidden color state $|CC\rangle_{str=-6,ST=00}$ is
relatively small. But for the case of d*,the energy of the hidden
color state in $(\Delta\Delta)_{ST=30}$ case are relatively not as
high as in the $\Omega\Omega$ case, and the cross matrix elements
between $(\Delta\Delta)_{ST=30}$ and its corresponding hidden
color state is relatively large. All of these features can be
understood based on the quite different flavor and spin structures
of $(\Omega\Omega)_{0^+}$ and d*.

\begin{table}
\caption{Coefficients of spin-flavor-color operators.}
\begin{small}
\begin {center}
\begin{tabular}{cccccccccc}
\hline\hline
  $\widehat{O}_{ij}$&&$\begin{array}{c}\Omega\Omega\\
\Omega\Omega\end{array}$&$\begin{array}{c}\Omega\Omega\\CC
\end{array}$&$\begin{array}{c}CC\\CC\end{array}$\\
\hline
&1&27&0&27\\
&$P_{36}$&-3&-12&-21\\
\hline
$\lambda_{i}^{c}\cdot\lambda_{j}^{c}$&$\widehat{O}_{12}$&-72&0&-18\\
&$\widehat{O}_{36}$&0&0&-36\\
&$\widehat{O}_{12}P_{36}$&8&32&2\\
&$\widehat{O}_{36}P_{36}$&-16&8&32\\
&$\widehat{O}_{13}P_{36}$&8&32&20\\
&$\widehat{O}_{16}P_{36}$&8&-4&20\\
&$\widehat{O}_{14}P_{36}$&-4&2&35\\
\hline
$(\vec{\sigma}_{i}\cdot\vec{\sigma}_{j})(\lambda_{i}^{c}\cdot\lambda_{j}^{c})$
&$\widehat{O}_{12}$&-72&0&-90\\
&$\widehat{O}_{36}$&0&-48&-60\\
&$\widehat{O}_{12}P_{36}$&8&32&74\\
&$\widehat{O}_{36}P_{36}$&112&-8&88\\
&$\widehat{O}_{13}P_{36}$&8&32&68\\
&$\widehat{O}_{16}P_{36}$&8&44&68\\
&$\widehat{O}_{14}P_{36}$&12&42&63\\
\hline
$(\vec{\sigma}_{i}\cdot\vec{\sigma}_{j})(\lambda_{8}^{f}(i)\cdot\lambda_{8}^{f}(j))$
&$\widehat{O}_{12}$&36&0&-36\\
&$\widehat{O}_{36}$&-60&0&-12\\
&$\widehat{O}_{12}P_{36}$&-4&-16&44\\
&$\widehat{O}_{36}P_{36}$&28&16&4\\
&$\widehat{O}_{13}P_{36}$&-4&-16&20\\
&$\widehat{O}_{16}P_{36}$&-4&32&20\\
&$\widehat{O}_{14}P_{36}$&12&24&0\\
\hline $\vec{\sigma}_{i}\cdot\vec{\sigma}_{j}$
&$\widehat{O}_{12}$&27&0&-27\\
&$\widehat{O}_{36}$&-45&0&-9\\
&$\widehat{O}_{12}P_{36}$&-3&-12&33\\
&$\widehat{O}_{36}P_{36}$&21&12&3\\
&$\widehat{O}_{13}P_{36}$&-3&-12&15\\
&$\widehat{O}_{16}P_{36}$&-3&24&15\\
&$\widehat{O}_{14}P_{36}$&9&18&0\\
 \hline
$\lambda_{8}^{f}(i)\cdot\lambda_{8}^{f}(j)$
&$\widehat{O}_{12}$&36&0&36\\
&$\widehat{O}_{36}$&36&0&36\\
&$\widehat{O}_{12}P_{36}$&-4&-16&-28\\
&$\widehat{O}_{36}P_{36}$&-4&-16&-28\\
&$\widehat{O}_{13}P_{36}$&-4&-16&-28\\
&$\widehat{O}_{16}P_{36}$&-4&-16&-28\\
&$\widehat{O}_{14}P_{36}$&-4&-16&-28\\
 factor&&$\frac{1}{27}$&$\frac{1}{27}$&$\frac{1}{27}$\\
\hline
\end{tabular}
\end{center}
\end{small}
\end{table}
\begin{table}
\caption{Binding energy $B$ and rms $\overline{R}$ of the
$(\Omega\Omega)_{0^+}$ dibaryon in coupled channel calculation.
$B=2M_{\Omega}-E_{(\Omega\Omega)_{0^+}}$,
$\overline{R}=\sqrt{\langle\,r^2 \rangle}$.}
\begin{small}
\begin {center}
\begin{tabular}{lllcc}
\hline\hline &&&$\Omega\Omega(L=0)$& $\begin{array}{c}\Omega\Omega\\
CC\end{array}(L=0)$\\
\hline
Chiral SU(3) &&$B$(MeV)&170.9&171.0\\
 quark model&&$\overline{R}$(fm)&0.62&0.62\\
 &&&&\\
Extended chiral &~ set I~ &$B$(MeV)&135.6&135.7\\
SU(3) quark model&&$\overline{R}$(fm)&0.60&0.60\\
&~ set II~ &$B$(MeV)&158.0&158.1\\
&&$\overline{R}$(fm)&0.59&0.59\\
\hline\hline
\end{tabular}
\end{center}
\end{small}
\end{table}

\subsection{ color screening effect}

It is well known that in the two-color-singlet-cluster system, the
form and the strength of the confining potential do not affect the
resultant quantities much. In concerning to the
$(\Omega\Omega)_{0^+}$ structure, the hidden-color channel $CC$ is
added to enlarge the model space. Once the $CC$ channel is
considered, the color Van der Waals force appears. To eliminate this
unreasonable force, one may use an error-function-like confining
potential to take the color screening effect, namely, the
nonperturbative QCD effect, into account~\cite{yuan}. Therefore, it
is necessary to examine the stability of the resultant binding
energy with respect to the form of the confining potential in the
presence of the hidden-color state. For this purpose, we also adopt
an error-function-like confining potential
\begin{equation}
V_{ij}^{\textrm{erf-conf}}=-(\lambda_{i}^{a}\lambda_{j}^{a})_{c}
\bigg[a^{0}_{ij}+a_{ij}\,\mathit{erf}\bigg(\frac{r}{l_{cs}}\bigg)\bigg]
\,\, ,
\end{equation}
where $l_{cs}$ denotes the color screening length, which is taken to
be 2.0 fm in the $(\Omega\Omega)_{0^+}$  structure calculation. The
results are shown in Table VI, from which one sees that the
resultant binding energy  of $(\Omega\Omega)_{0^+}$ are quite
similar to those in the quadratic confinement case, namely, the
bound state property would not change much when the color screening
effect is counted.

\begin{table}
\caption{Binding energy $B$ of the $(\Omega\Omega)_{0^+}$ dibaryon
with $r^2$ and with error-function-like confinements in single
channel calculation. $B=2M_{\Omega}-E_{(\Omega\Omega)_{0^+}}$.}
\begin{small}
\begin {center}
\begin{tabular}{lllccc}
\hline\hline &&&$r^2$ && Erf\\
 \hline
Chiral SU(3)quark model &&$B$(MeV)&170.9&&163.6\\
&&$\overline{R}$(fm)&0.62&&0.62\\
 &&\\
Extended chiral & set I~ &$B$(MeV)&135.6&&126.1\\
&&$\overline{R}$(fm)&0.60&&0.61\\
SU(3) quark model& set II~ &$B$(MeV)&158.0&&149.1\\
&&$\overline{R}$(fm)&0.59&&0.59\\
\hline\hline
\end{tabular}
\end{center}
\end{small}
\end{table}

\subsection{The mixing of scalar mesons }

\begin{table}
\caption{Binding energy $B$ and rms $\overline{R}$ of the
$(\Omega\Omega)_{0^+}$ dibaryon  in a coupled channel calculation ~
$B=2M_{\Omega}-E_{\textrm{$(\Omega\Omega)_{0^{+}}$}}$,
$\overline{R}=\sqrt{\langle\,r^2 \rangle}$, the parameters taken
from Ref.\cite{huang1}.}
\begin{small}
\begin {center}
\begin{tabular}{lllccccc}
\hline &&&$B$(MeV)&$\overline{R}$(fm) \\ \hline
Chiral SU(3) &~ set I ($\theta^{S}=35.264$)~ &&61.12&0.70\\
quark model&~ set II ($\theta^{S}=-18$)~~ &&134.5&0.64\\
&&&&\\
Extended chiral&~ set I ($\theta^{S}=35.264$) ~&&15.9&0.85\\
 SU(3) quark model&~ set II ($\theta^{S}=-18$)&&75.6&0.65\\
\hline\hline
\end{tabular}
\end{center}
\end{small}
\end{table}

The chiral SU(3) quark model has been widely used to study the
baryon-meson interaction. When the model is extended to study the
kaon-nucleon (KN) scattering, the scalar meson mixing between the
flavor singlet and octet mesons is considered to explain the
experimental phase shift. Therefore, the effect of the scalar meson
mixing is also studied. In our calculation, scalar $\sigma$,
$\epsilon$ mesons are mixed from $\sigma_{0}$ and $\sigma_{8}$ with
\begin{eqnarray}
\sigma =\sigma_{8}\sin\theta^{S}+\sigma_{0}\cos\theta^{S},\nonumber\\
\epsilon~ =\sigma_{8}\cos\theta^{S}-\sigma_{0}\sin\theta^{S},
\end{eqnarray}
The mixing angle $\theta^{S}$ has been an open issue because the
structure of the $\sigma$ meson is still unclear and controversial.
Here we adopt two possible values as did in \cite {huang,huang1}.
One is the ideal mixing with $\theta^{S}= 35.264^\circ$. This is an
extreme case in which the $\sigma$ exchange may occur only between
u(d) quarks, while $\epsilon$ occurs between s quarks. Another
mixing angle with $\theta^{S}= -18^\circ$ adopted was provided by
Dai and Wu based on their investigation of a dynamically spontaneous
symmetry breaking mechanism \cite{daiyb}. The calculated results of
$\Omega\Omega_{(0^{+})}$ are shown in Table VII, of which the
parameters are taken from Ref.\cite{huang1} by fitting KN scattering
processes.

From the results one sees that in the extended chiral SU(3) quark
model, the binding energy of $\Omega\Omega_{(0^{+})}$ has very big
difference ( from 134 MeV to 15 MeV ) for different mixing angle of
scalar mesons. When the mixing of scalar meson is taken to be the
ideally mixing, that means there is no $ \sigma$ exchange between
two $s$ quarks, the attraction from the $\sigma$ meson has reduced
to zero, and thus the binding energy becomes much smaller. Comparing
with the case of the chiral SU(3) quark model in which the binding
energy of $\Omega\Omega_{(0^{+})}$ is about 60 MeV even the mixing
angle is taken as ideal mixing, this means that in the extended
chiral SU(3) quark model the short range repulsive interaction
offered by $\phi$ meson exchange is stronger than that of OGE in the
chiral SU(3) quark model. At the same time, in the case of
$\theta^{S}= -18^\circ$, the attraction of the $\sigma$ meson is
still play important role, and thus the binding energy becomes
larger. In this sense, all of these features are useful for
examining the model and its corresponding parameters.

Here we would like to point out that when the mixing angle of
scalar meson is considered, the parameters are obtained by fitting
KN scattering processes, not by fitting NN and YN scattering
processes. At the moment, we could not get a set of unified
parameter to fit all of the scattering data of KN, NN and YN
systems. However, we would like to see the effect on the structure
of $\Omega\Omega_{(0^{+})}$ from various set of parameters. The
results tell us that different sets of parameters have large
effect on the structure of $\Omega\Omega_{(0^{+})}$ dibaryon,
especially for the extreme case of the ideal mixing. From the
above discussion, one can see that no matter which set of
parameters is taken, even in the case of scalar meson ideal mixing
the attraction from $\epsilon$ is almost cancelled by the
repulsion from $\phi$ exchange, the $\Omega\Omega_{(0^{+})}$ is
always a bound state. This is because its symmetry structure is
very special \cite{zhprc61}, and the quark exchange effect of
$\Omega\Omega_{(0^{+})}$ is really important to make it always
bound. For other systems without such special symmetry structure,
the binding energies would also changed a lot with different sets
of parameters. For some weakly bound systems, the structure
property could be changed  from bound state to unbound state.
There would be no such stable bound property like
$\Omega\Omega_{(0^{+})}$ in the systems without such special
symmetry property.

\section{summary}
In this work, the structure of $\Omega\Omega_{(0^{+})}$ dibaryon
with strangeness $s=-6$ is studied in the extended chiral $SU(3)$
quark model.  The model space is enlarged by including the CC
channel. The calculations are performed by solving a
coupled-channel RGM equation. Firstly, the vector meson exchange
effect on $(\Omega\Omega)_{0^+}$ dibaryon is studied on quark
level. The results show that $(\Omega\Omega)_{0^+}$ is still a
deeply bound state when the vector meson exchanges control the
short range part of the quark-quark interaction, which is quite
similar to the results obtained from the chiral $SU(3)$ quark
model. Secondly, the effect from hidden color channel on
$\Omega\Omega_{(0^{+})}$ dibaryon is studied.  It is found that
the energy of the hidden color state $|CC\rangle_{str=-6,ST=00}$
is much higher than that of $(\Omega\Omega)_{0^+}$ state, and the
cross matrix elements between $(\Omega\Omega)_{0^+}$ and its
corresponding hidden color state $|CC\rangle_{str=-6,ST=00}$ is
relatively small, which explains why the CC channel has little
effect on the binding energy of $\Omega\Omega_{(0^{+})}$ dibaryon
in contrast to the deltaron dibaryon case with a different
situation. In addition, the error function confinement potential
is considered, from which the resultant binding energy of
$(\Omega\Omega)_{0^+}$ is quite similar to that in the quadratic
confinement case. Namely, the bound state property would not
change much when the color screening effect is counted.  Finally,
the effects of scalar meson mixing on $\Omega\Omega_{(0^{+})}$
dibaryon are also investigated. The result shows that the binding
energy of the $\Omega\Omega_{(0^{+})}$ has changed substantially
with different scalar meson mixings, which becomes larger with
$\theta^{S}= -18^\circ$ and smaller in the ideal mixing with
$\theta^{S}= 35.264^\circ$. It should be noted that our current
analysis of $\Omega\Omega_{(0^{+})}$ dibaryon is based on the
results from the fit to KN scattering processes, in which the
scalar meson mixing angle should be considered. A further study of
the $\Omega\Omega_{(0^{+})}$ dibaryon structure should be based on
results obtained from a fit to KN, NN and YN scattering processes
simultaneously, at least from a fit to NN and YN scattering
processes, which will be considered in the near future.

Our conclusion is that the $\Omega\Omega_{(0^{+})}$ dibaryon is
always a bound state regardless which set of parameters is used in
the extended chiral $SU(3)$ quark model due to its unique symmetry
property. Therefore, the $\Omega\Omega_{(0^{+})}$  is the most
interesting dibaryon candidate.

Supported from the National Natural Science Foundation of China
(10475087; 10575047) is acknowledged.


{}

\end{document}